\documentclass[aps,prl]{revtex4}

\usepackage{graphicx}% Include figure files

\usepackage{epstopdf}	% this graphic package converts eps to pdf for easier usage of eps files

\usepackage{lpic}
\usepackage{amsmath,amssymb}
\usepackage{nicefrac}
\usepackage[utf8]{inputenc}
\usepackage{hyperref}
\hypersetup{
    colorlinks=false,       % false: boxed links; true: colored links
    colorlinks=true,
    linkcolor=blue,          % color of internal links (change box color with linkbordercolor)
    citecolor=red,        % color of links to bibliography
    filecolor=magenta,      % color of file links
    urlcolor=cyan           % color of external links
}
\usepackage{xcolor}
\usepackage{listingsutf8}
\newcommand{\ket}[1]{\left|{#1}\right\rangle}

\begin{document}

\title[Coherent Control]{Coherent control of solid state nuclear spin nano-ensembles}

\author{Thomas Unden$^{1}$, Nikolas Tomek$^{1}$, Timo Weggler$^{1}$, Florian Frank$^1$, Paz London$^{2}$, Jonathan Zopes$^{3}$, Christian Degen$^{3}$, Nicole Raatz$^{4}$, Jan Meijer$^{4}$, Hideyuki Watanabe$^{5}$, Kohei M. Itoh$^{6}$, Martin B. Plenio$^{7}$, Boris Naydenov$^{1}$\footnote{E-mail: boris.naydenov@uni-ulm.de} and Fedor Jelezko$^{1}$}
\affiliation{$^1$Institute for Quantum Optics and Center for Integrated Quantum Science and Technology (IQST), Albert-Einstein-Allee 11, Universit\"at Ulm, 89069 Ulm, Germany}
\affiliation{$^2$Department of Physics, Technion, Israel Institute of Technology, Haifa 32000, Israel}
\affiliation{$^{3}$ Department of Physics, ETH Zurich, Otto Stern Weg 1, 8093, Zurich, Switzerland.}
\affiliation{$^{4}$ Felix Bloch Institute for Solid State Physics, Universität Leipzig, 04103 Leipzig, Germany.}
\affiliation{$^5$ Correlated Electronics Group, Electronics and Photonics Research Institute, National Institute of Advanced Industrial Science and Technology (AIST), Tsukuba, Japan}
\affiliation{$^6$ Department of Applied Physics and Physico-Informatics, Faculty of Science and Technology, Keio University, Yokohama, Japan}
\affiliation{$^7$ Institute for Theoretical Physics and Center for Integrated Quantum Science and Technology (IQST), Albert-Einstein-Allee 11, Universit\"at Ulm, 89069 Ulm, Germany}
%\ead{boris.naydenov@uni-ulm.de}

%\renewcommand{\comment}[1]{{\it \color{red} #1 }}

\begin{abstract}
Detecting and controlling nuclear spin nano-ensembles is crucial for the further development of nuclear magnetic resonance (NMR) spectroscopy and for the emerging solid state quantum technology. Here we present the fabrication of a $\approx$ 1 nanometre thick diamond layer consisting of $^{13}$C nuclear spins doped with Nitrogen-Vacancy centres (NV) embedded in a spin-free $^{12}$C crystal matrix. A single NV in the vicinity of the layer is used for polarization of the $^{13}$C spins and the readout of their magnetization. We demonstrate a method for coherent control of few tens of nuclear spins by using radio frequency pulses and show the basic coherent control experiments - Rabi oscillations, Ramsey spectroscopy and Hahn echo, though any NMR pulse sequence can be implemented. The results shown present a first steps towards the realization of a nuclear spin based quantum simulator.

\end{abstract}

\maketitle

The concept of a quantum simulator originates from Feynman \cite{Feynman82}, where the idea is to use a well controlled quantum system to simulate different types of Hamiltonians. While the first demonstrations of quantum simulator concepts have been realized already in ultra cold quantum gases \cite{Bloch12} and ion traps~\cite{Blatt16}, but a solid state implementation is still limited, though there has been some promising realizations of interaction Hamiltoninians using macroscopic nuclear spin ensembles \cite{Suter15} (not scalable though due to the usage of pseudo-pure states) and using superconducting qubits \cite{Roushan17}. A recent theoretical proposal and analysis demonstrate that a quantum simulator even for 2D spin systems is feasible on the basis of diamond quantum technologies.

Two major challenges towards this goal are the fabrication of a nano-ensemble of coupled nuclear spins and their coherent control and read-out. Important steps towards the latter goal have been taken during the last decades as the minimum number of  nuclear spins that can be detected has been continuously decreasing, where finally coherent control of a single nuclear spin strongly coupled to a single electron spin has been demonstrated using optical \cite{Jelezko04b} and electrical detection \cite{Morello13}. Later on first the detection of ten thousand nuclear spins outside of the substrate has been demonstrated \cite{Staudacher13, Mamin13} and finally single spin sensitivity \cite{Mueller14, Sushkov14, Du14} has been achieved.

Here, we address the above challenges by proposing a novel approach for generating clusters of nuclear spins, with intra-clusters interactions of varying strength. Additionally we demonstrate a method for initialization, readout and control of few tens of nuclear spins. We fabricated by chemical vapour deposition (CVD) a nanometre thick diamond layer of $^{13}$C carbon atoms (nuclear spin $I=1/2$) on two substrates, referred to as samples A and B. The growth conditions and procedure have been reported previously \cite{Watanabe11}. In the former the layer is separated both from the substrate and from the surface by a 10 nm thick $^{12}$C enriched (nuclear spin free) diamond layer in order to reduce magnetic noise. In sample B the distances are 20 and 5 nm respectively. In sample A the $^{13}$C layer was additionally doped with nitrogen and sample B was implanted with nitrogen ions, in order to create single nitrogen-vacancy centres (NV) in the vicinity of the $^{13}$C layer. In both samples single NVs are coupled to few tens of nuclear spins, thus enabling polarization and read out of the magnetization of these small ensembles. Coherent control over the nuclear spins (demonstrated in sample A) is realized via radio frequency (RF) pulses allowing to perform NMR spectroscopy as well as to implement quantum gates.

\begin{figure}[htb]
\centering
\includegraphics[width=0.4\textwidth]{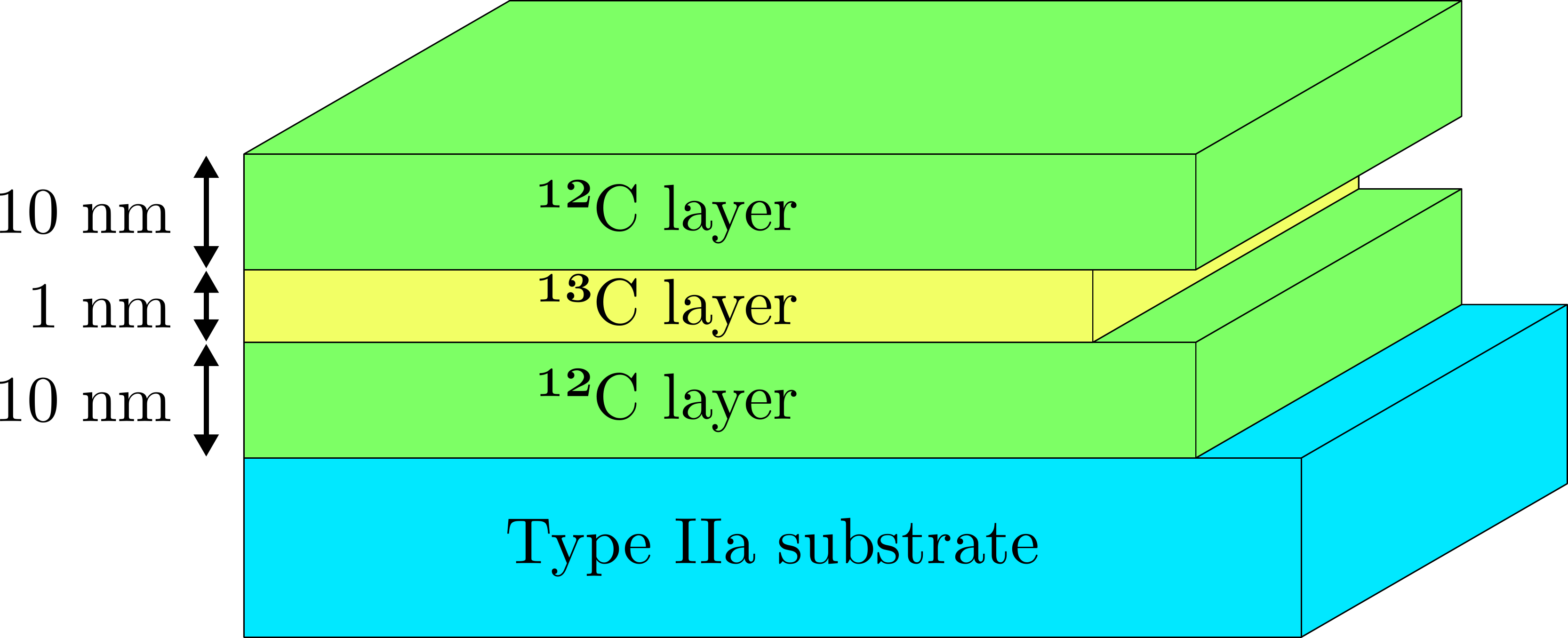}
\caption[Schematic drawing]{Schematic drawing of a cross section of sample A. In sample B the first $^{12}$C layer is 20 nm and the cap layer 5 nm thick.}
\label{FigSample}
\end{figure}

A cross section of the diamond samples used in the experiments is shown in Figure~\ref{FigSample}.
First a $^{12}$C enriched (99.99 \%) diamond layer is grown via CVD on an ultra pure diamond substrate (E6 Ltd., electronic grade). On top of it a $^{13}$C (using $>98.4$ \% $^{13}$CH$_4$) enriched layer is grown which in sample A is additionally doped with nitrogen in order to form single NV centres. Finally the cap layer is used to protect the spin properties of the NV centres from the surface noise which induces decoherence as reported previously \cite{Staudacher12}.

\begin{figure*}[h!bt]
\centering

\includegraphics[scale=0.7]{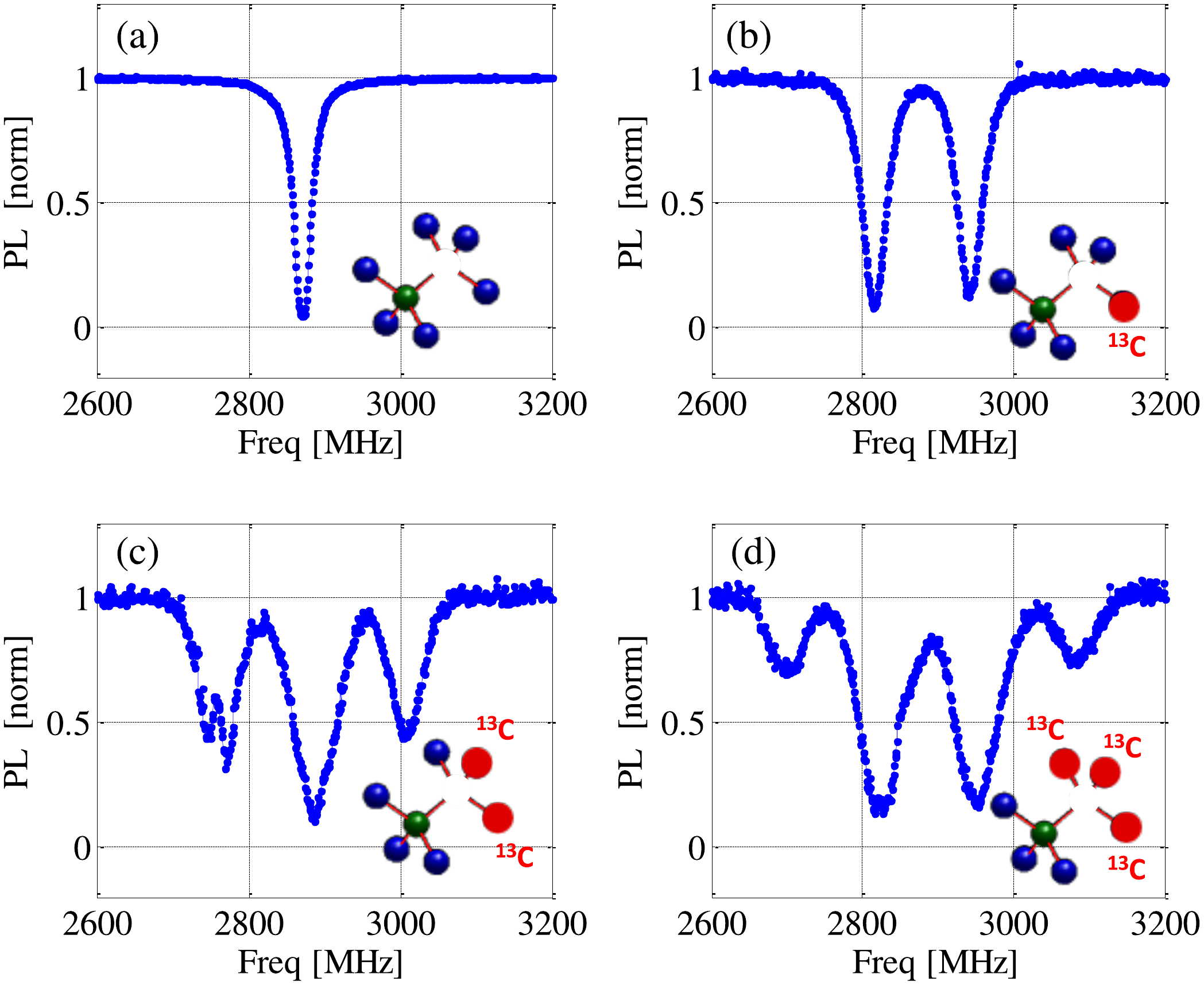}

\protect\caption[ODMR Spectra]{ODMR spectra of single NV-centres at zero magnetic field measured in the overgrown diamond layer having none (a), one (b), two (c) and three (d) $^{13}$C carbon atoms in the first shell.}
\label{FigODMR} 
\end{figure*}

Nitrogen $\delta$-doping during the growth process (sample A) and the nitrogen ion implantation (sample B) were optimized to position the NV-centres in the vicinity of the $^{13}$C enriched region. Three implantation energies were used - 5, 2.5 and 1 keV, resulting in an average depth of the nitrogen ions of 7, 3.5 and 1.4 nm respectively. The nuclear spin structure can be characterized by measuring the effect of the nuclear spin environment on the NV-centres. Confocal microscopy fluorescence imaging revealed the presence of single NVs in both samples. Optically detected magnetic resonance measurements of over 483 NV centres (sample A) and 584 NV centres (sample B) show strong coupling of the NVs to $^{13}$C nuclear spins. The spectra can be divided into four groups, distinguished by the number of $^{13}$C nuclear spins next to the vacancy, see Figure~\ref{FigODMR}.

NV-centres (A) lacking a first-shell carbon spin (Fig.\ref{FigODMR}a), (B) interacting with a single first-shell $^{13}$C spin, and showing a 130 MHz characteristic splitting (Fig.\ref{FigODMR}b), (C) interacting with two first-shell $^{13}$C spins showing three spectral lines (Fig.\ref{FigODMR}c) and (D) interacting with three first-shell $^{13}$C spins showing four spectral lines (Fig.\ref{FigODMR}d). The latter group would dominate the observed spectra if all the NV-centres were embedded in a 100\% abundance of $^{13}$C atoms. We note the probability of finding an NV-center of group $k$ ($k=A,B,C,D$) as $p_{k},$ and summarize the values obtained from the data for both samples in table~\ref{TableODMR}.
\begin{table*}[htb]
\begin{tabular}{|c|c|c|c|c|c|}
\hline
Probability in \% & Sample A & \multicolumn{3}{|c|}{Sample B} & Theoretical model\\
\hline
 \multicolumn{2}{|c|}{ } & $E_{\mathrm{impl.}}=5$ keV & $E_{\mathrm{impl.}}=2.5$ keV & $E_{\mathrm{impl.}}=1$ keV & \\
\hline
$p_A$ & 73.4 & 88.1 & 84.5 & 88.5 & 72.75\\
\hline
$p_B$ & 13.7 & 11.9 & 14.6 & 11.5 & 13.065\\
\hline
$p_C$ & 7.2 & 0 & 1.0 & 0 &  9.075\\
\hline
$p_D$ & 5.7 & 0 & 0 & 0 & 5.11\\
\hline
\end{tabular}
\caption{Probability of finding the four groups of NV centres in samples A and B.}
\label{TableODMR}
\end{table*}

%and find \begin{equation} 
%p_{A}=73.4\%,p_{B}=13.7\%,p_{C}=7.2\%,p_{D}=5.7\%.
%\end{equation}
We note the larger number of NVs in groups $p_{A}$ and $p_{D}$ suggest the presence of two types of regions - one with high and one with low concentration of $^{13}$C spins at the microscopic level.

To quantify the overgrown layer at the nanometre scale, we fit a model whose principles are derived from the growth procedure. A detailed description is given in the Supplementary Material.
\begin{figure}
\centering
\includegraphics[scale=0.4]{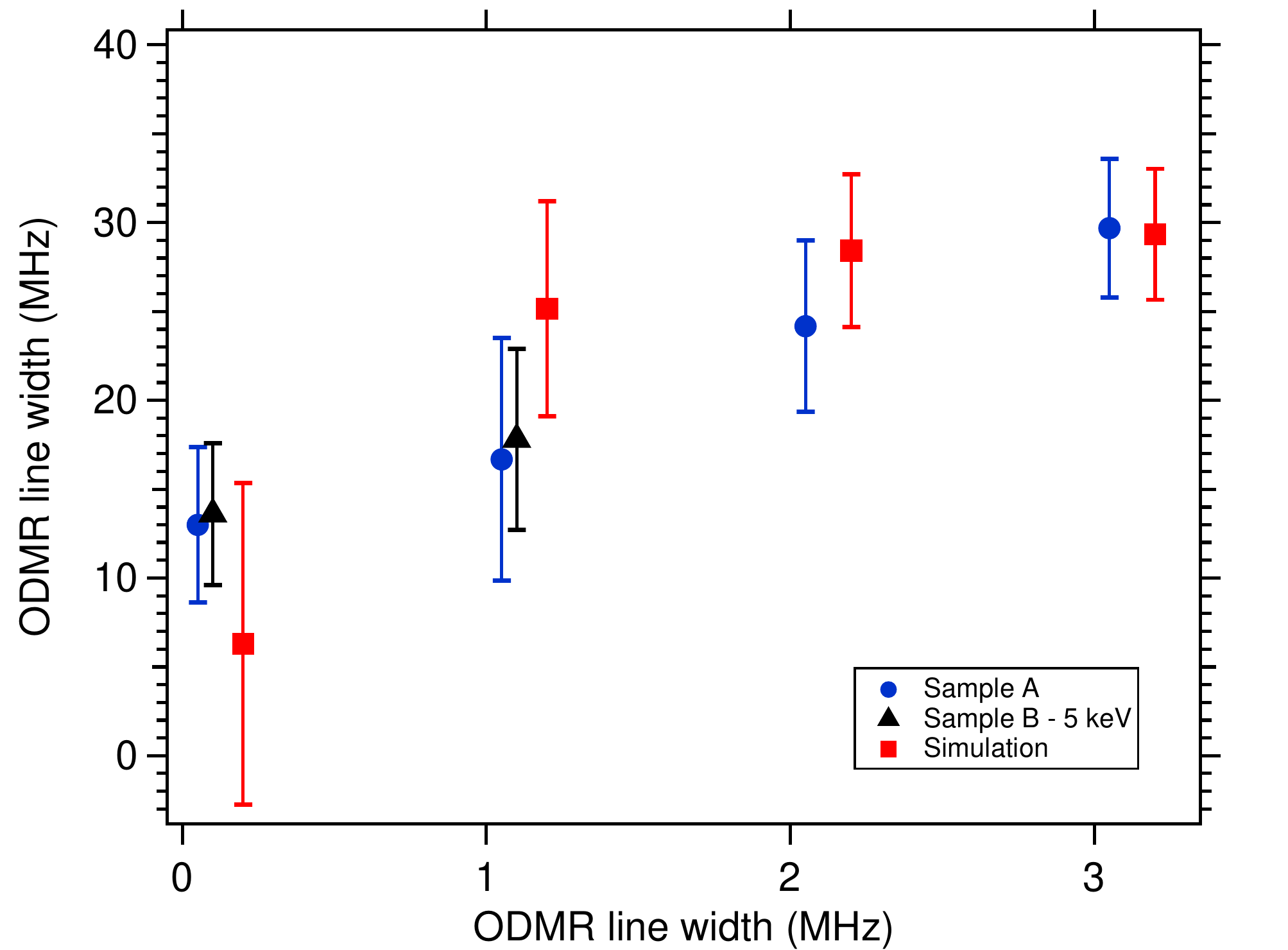}

\protect\caption{Quantitative characterization of the $^{13}$C layer. The averaged ODMR for NV-centres in group A-D (blue markers). The error bars represents one standard deviation in the width of measured spectra for group A-D. The red markers are simulation with the optimal parameters $(d/Z)=0.39$ and $(\lambda/Z)=0.2$ See text and the Supplementary Material for more details.}
\label{FigODMRFit}
\end{figure}

For each of the four groups, we extract the averaged line width of the ODMR spectra, and compare it with the one expected from simulation (data not shown). In the simulated spectra, the interaction between the electron  and the nuclear spins is extracted from an exact calculation of the hyperfine tensor \cite{Nisovtsev14, Nisovtsev17} for the nuclear spins which are located within 1.5 nm from the NV-center. For nuclear spins which are further away, the dipole-dipole approximation is used. The line width $\Delta\nu$ of the NV-centres increases with the number of first-shell $^{13}$C spins, and the distribution of $\Delta\nu$ for the group D (three nuclear spins) is smaller than the one of group B and C. Both features are reproduced by our model in a good agreement (Fig.\ref{FigODMRFit}) with the experimental data. For additional analysis of the structure of the layer, see the Supplementary Material.

NV centres from group A in sample A were chosen for the further experiments since in this diamond we found single NV coupled to a larger number of $^{13}$C spins compared to sample B. The dense solid state $^{13}$C nuclear spin ensemble, coupled to NVs from group A, can be used to explore many body physics and as a test system for improving the resolution and sensitivity of nanoscale NMR techniques. A single NV from group A was chosen, having parallel and perpendicular components of the hyperfine interaction to $^{13}$C nuclear spins of $A_{\parallel}\sim A_{\perp}\sim50$~kHz \cite{Taminiau12}, measured both by XY8 \cite{Taminiau12} and Hartmann-Hahn \cite{London13} type of measurements. From this the distance between the NV and the $^{13}$C layer is estimated to be 0.72 nm. The coherence time of this NV was found to be $T_2=50\;\mu$s.

\begin{figure*}[bht]
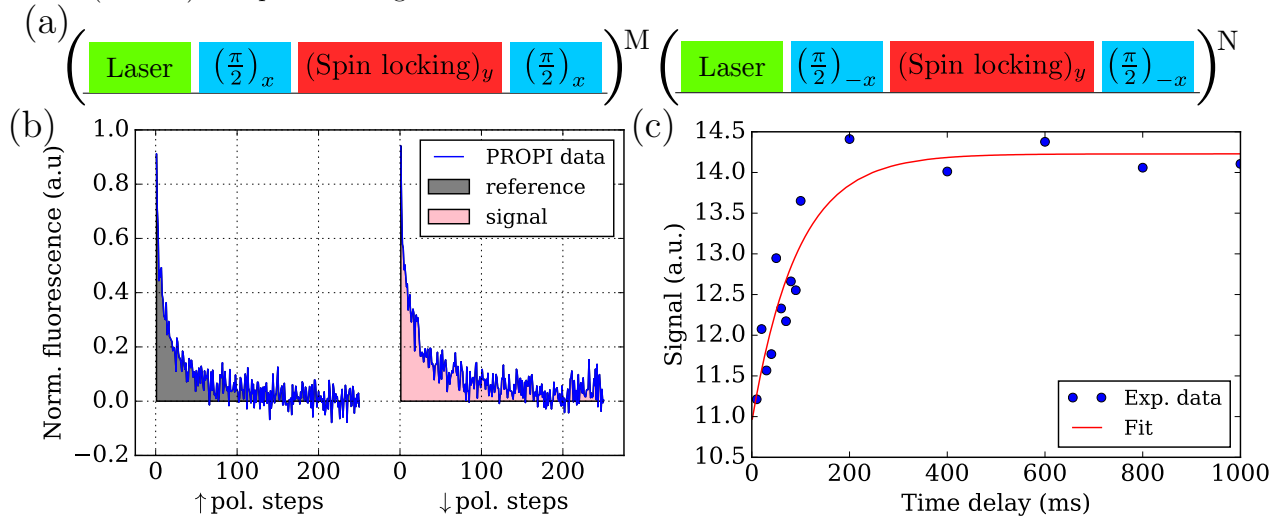

\centering
\begin{lpic}{Figure4a(0.45,0.45)}
\lbl[t]{-6,30;  \Large{(a)}}
\end{lpic} 

\begin{tabular}{cc}
\begin{lpic}[l(0mm)]{Figure4b(0.55,0.55)}
\lbl[t]{0,101;  \Large{(b)}} 
\end{lpic} &  
\begin{lpic}{Figure4c(0.55,0.55)}
\lbl[t]{0,101;  \Large{(c)}}
\end{lpic} 
\end{tabular}
\caption[PROPI]{(a) PROPI pulse sequences. In all experiments reported here $M=N=100$ and the length of the spin locking pulse was set to 20~$\mu$s. (b) NV fluorescence as a function of the number of polarization steps. (c) Spin lattice relaxation time $T_1=100$ ms of a small nuclear spin ensemble. }
\label{FigPROPI}
\end{figure*}

In order to obtain control over the nuclear spins, a robust method for the initialization (polarization) and readout of the nuclear spins is required. We have recently developed a technique to achieve this goal by utilizing a Hartmann-Hahn double resonance \cite{Scheuer17} and here the main idea is given. The NV's electron spin is driven by applying a spin locking sequence with a Rabi frequency $\Omega=\omega_L$, where $\omega_L$ is the Larmor frequency of the nuclear spins. At this condition there is transfer of polarization between the two systems, allowing to polarize and read out the state of a small ensemble of nuclear spins. The pulse sequence of this measurement named Polarization ReadOut via Polarization Inversion (PROPI) is depicted in Figure~\ref{FigPROPI}.

It consists of two blocks of pulses and the working principle is the following. First we apply a laser pulse to initialize the NV center into the $\ket{m_s=0}$ state. Then we apply a MW $\pi/2$ pulse to create the superposition state $1/\sqrt{2}\left(\ket{m_s=0}+\ket{m_s=-1}\right)$ followed by a long MW pulse to keep the NV in this rotated basis (spin locking) to allow for resonant transfer of population between NV and nuclei. Afterwards a $\pi/2$ pulse is used to transfer the coherence to population difference, which is then read out by a second laser pulse. This sequence is repeated several hundred times and the signal is shown in Figure~\ref{FigPROPI}b. We observe that with increasing number of steps, the fluorescence decays since the NV's electron spin polarization is transferred to the surrounding nuclear spins. After some time a saturation is reached, where the nuclear spins located in the vicinity of the NV center are polarized parallel to the applied static magnetic field $B_0$ ("up" state $\ket{\uparrow}$ \cite{Scheuer17}). Now if we change the phase of the first MW pulse by 180$^{\circ}$ in order to populate the opposite dressed state and we repeat the sequence, we observe a similar behaviour as there is again spin polarization transfer. The difference is that here the nuclear spin polarization is changed from parallel to anti-parallel to the magnetic field ("down" state $\ket{\downarrow}$). If the nuclear spin polarization has not changed in between the two pulse blocks, then the area $A_{\uparrow}$ below the first curves will be equal to the area $A_{\downarrow}$ below the second curve. However, if there is a process affecting the nuclear spins (thermal relaxation, decoherence and spin diffusion, see below), then $A_{\uparrow}\neq A_{\downarrow}$. The sequence could be simplified if we remove the second pulse block and polarize the nuclear spins only in one state, for example in $\ket{\uparrow}$. In this case the area below the curve will change due to some nuclear spin dynamics, for example if there is a time delay $\tau$ then $A_{\uparrow}\neq A_{\uparrow}(\tau)$. Both the PROPI sequence (Figure~\ref{FigPROPI}a) and its simplified version can be used to initialize the nuclear spin ensemble and readout its magnetization, where we obtain similar results. First we demonstrate that the spin lattice relaxation (SLR) time $T_1$ of the nuclear spins can be determined by using the simplified PROPI. For this purpose we introduce a time delay $\tau^{\mathrm{SLR}}$ between the two polarization blocks, see Figure~\ref{FigPROPI}b. Here we observe that $A_{\uparrow}(\tau^{\mathrm{SLR}})$ increases with increasing $\tau^{\mathrm{SLR}}$ and a saturation is reached around 200~ms. An exponential decay fit reveals $T_1=100$~ms, a quite short value which is probably due to the low magnetic field $B_0=458$~G and the high local concentration of paramagnetic subsitutional nitrogen (P1 centres, electron spin $S=1/2$). Another possible loss of polarization could be the nuclear spin diffusion out of the ensemble measured by the NV center.

\begin{figure*}[bht]
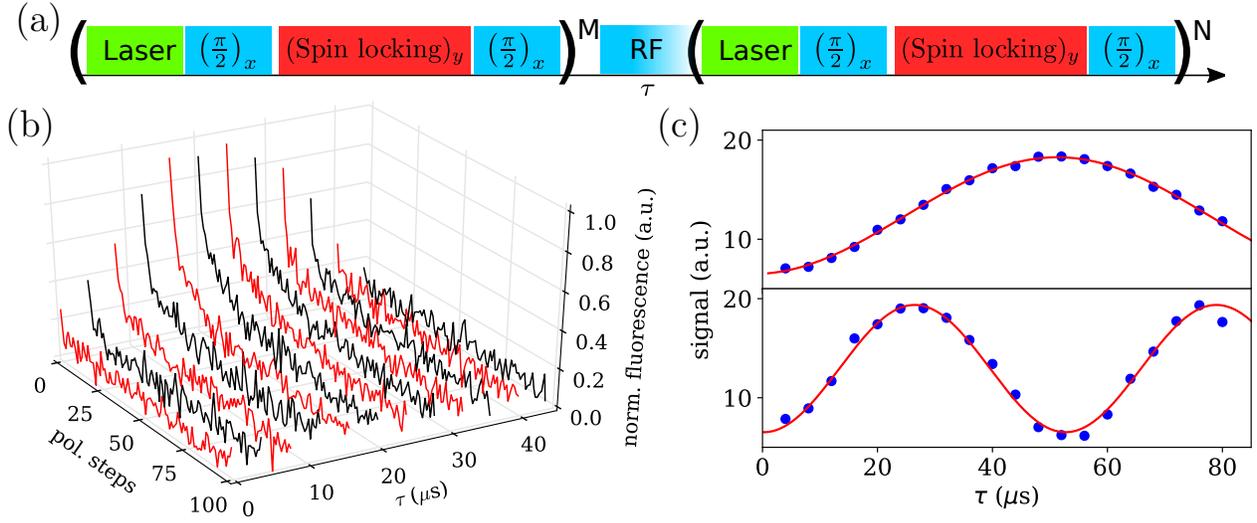

\centering
\begin{lpic}{Figure5a(0.425,0.425)}
\lbl[t]{-8,30;  \Large{(a)}}
\end{lpic} 

\begin{tabular}{cc}
\begin{lpic}[l(0mm)]{Figure5b(0.55,0.55)}
\lbl[t]{0,100;  \Large{(b)}} 
\end{lpic} &  
\begin{lpic}{Figure5c(0.55,0.55)}
\lbl[t]{0,100;  \Large{(c)}}
\end{lpic} 
\end{tabular}
\caption[C13Rabi]{(a) Pulse sequence for performing nuclear spin Rabi experiments ($M=N=200$). Raw (b) and processed data (c) from the PROPI measurement. In (c) the first period of Rabi oscillations is observed, where the frequency increase with increasing the MW power (upper to lower plot).}
\label{FigC13Rabi}
\end{figure*}

Next we demonstrate coherent control over the nuclear spins using RF pulses. In figure~\ref{FigC13Rabi}a the pulse sequence is shown used to measure Rabi oscillations of the nuclear spin ensemble.
Similar to the previous experiment it starts with a pulse block (see also figure~\ref{FigPROPI}a) to initialize the nuclear spins into the "up" state $\ket{\uparrow}$. Afterwards we apply a single RF pulse resonant with the $\ket{\uparrow}\leftrightarrow\ket{\downarrow}$ transition. Then we apply a second pulse block, which polarizes the nuclear spin ensemble back into the $\ket{\uparrow}$ state. The NV's fluorescence read out from this pulse block is depicted in figure~\ref{FigC13Rabi}b. We find that the signal increases with increasing the length of the RF pulse $\tau^{\mathrm{RF}}$. If we take the area below the curves, we obtain the typical Rabi oscillations, shown in Figure~\ref{FigC13Rabi}c, upper graph. By increasing the RF power we observe an increase in the Rabi frequency as expected, Figure~\ref{FigC13Rabi}c, lower graph.

\begin{figure*}[bht]
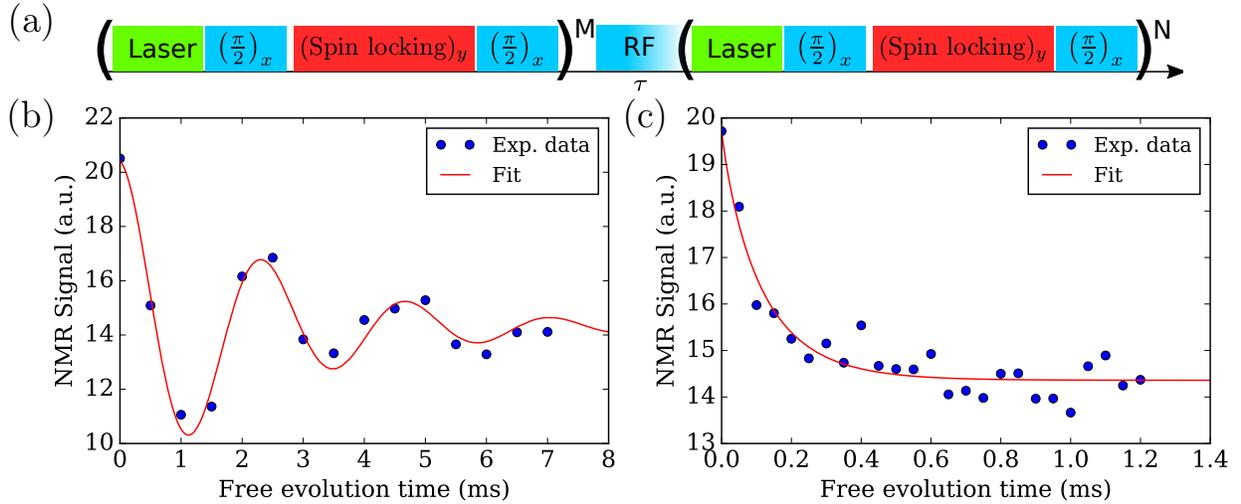

\centering
\begin{lpic}{Figure6a(0.4,0.4)}
\lbl[t]{-20,30;  \Large{(a)}}
\end{lpic} 

\begin{tabular}{cc}
\begin{lpic}[l(0mm)]{Figure6b(0.55,0.55)}
\lbl[t]{-3,100;  \Large{(b)}} 
\end{lpic} &  
\begin{lpic}{Figure6c(0.55,0.55)}
\lbl[t]{0,100;  \Large{(c)}}
\end{lpic} 
\end{tabular}
\caption[C13FID]{(a) Pulse sequence for measuring FID of the nuclear spins ($M=N=200$). FID  of the nuclear spin ensemble when the NV center is in the $\ket{m_s=0}$ (b) and in the $\ket{m_s=-1}$ (c) state. A shortening of the $T_2^*$ from 2.5 to 0.113~ms is observed.}
\label{FigC13FID}
\end{figure*}

Finally we demonstrate NMR spectroscopy of a small nuclear spin ensemble. For this purpose we record the free induction decay (FID) of the ensemble (sometimes called Ramsey fringes) by using the pulse sequence depicted in figure~\ref{FigC13FID}a.
After polarizing the nuclear spins into the $\ket{\uparrow}$ state we apply a RF $\pi/2$ pulse to create the superposition state $\ket{\psi}=1/\sqrt{2}\left(\ket{\uparrow}+\ket{\downarrow}\right)$. We let this state evolve for a time $\tau^{\mathrm{FID}}$ then we apply a second RF $\pi/2$ to convert it to a population difference. The latter is read out by the second pulse block on the NV center. In figure~\ref{FigC13FID}b we show a typical FID of a $^{13}$C nuclear spin ensemble, when the NV was initialized in $m_s=\ket{0}$ before the RF pulse. The exponential fit to the data reveals a phase memory time of $T_2^*=2.5$~ms which is probably limited by the spin relaxation time $T_1$ of the NV or by the high local concentration of P1 centres. If the NV is initialised into $\ket{m_s=-1}$ state by applying a laser and a MW $\pi$ pulse, then we observe a much shorter $T_2^*=0.113$~ms (figure~\ref{FigC13FID}c). This result can be explained by the fact, the nuclear spins experience a magnetic field gradient of about 33 G/nm (at 0.72 nm distance), generated by the electron spin of the NV. The gradient shifts the resonance frequencies of the nearby nuclear spins and a broadening of the line of about $33\times1.1$~kHz/G$=36.3\,$ kHz is expected, which agrees roughly with the value obtained from the experiment - $1/\pi T_2^*=1/(0.113\times\pi)=2.82$~kHz.

In conclusion, we have demonstrated the fabrication of a nanometre thin $^{13}$C enriched diamond layer, where small nuclear spin ensembles are coupled to single NV centres. We present a method for coherent control of those ensembles by combining RF pulses and PROPI-based \cite{Scheuer17} pulse sequences. Rabi measurements and NMR spectroscopy have been performed, revealing that the NMR line width depends strongly on the state of the NV center during the free evolution time. This result confirms that the $^{13}$C nuclear spins are indeed in the close vicinity of single NV centres as expected from the CVD growth conditions. We believe that our work will find application in the emerging field of solid state quantum simulators, based not only on NV centres, but also on other physical systems involving a central electron spin and a nuclear spin bath - such as phosphor donors in silicon and semiconductor quantum dots.
%\newpage

\section*{Acknowledgements}

This work was supported by the ERC Synergy grant BioQ, the EU (EQuaM, HYPERDIAMOND grant agreement No 667192), the DFG (SFB TR/21, FOR 1493) and the Volkswagenstiftung. BN is grateful to the Postdoc Network program of the IQST and to the Bundesministerium f\"ur Bildung und Forschung for receiving the ARCHES award.
The work at ETH Zurich was supported by Swiss National Science Foundation (SNFS) Project Grant No. 200020\underline{~~}175600, the National Center of Competence in Research in Quantum Science and Technology (NCCR QSIT), and the DIAmond Devices Enabled Metrology and Sensing (DIADEMS) program, Grant No. 611143, of the European Commission.
The work at Keio was supported in parts by KAKENHI (S) No. 26220602, SPS Core-to-Core Program, and Spin-RNJ
 
\section*{References}
 
%\bibliography{C13_Layer}
%\bibliographystyle{apsrev4-1}

%

\end{document}